\documentclass[12pt]{article}
\usepackage[utf8x]{inputenc}
\usepackage[T1]{fontenc}

\usepackage{hfstyle}

\usepackage{graphicx}
\usepackage{multirow}
\usepackage[all,cmtip]{xy}
\usepackage{amsmath, amssymb,bbm,color}

\newtheorem{prop}{Proposition}[section]
\newtheorem{cor}{Corollary}[section]

\newtheorem{thm}{Theorem}[section]

\begin{document}
% 
%\begin{frontmatter}
\title{Minimal entropy approximation for cellular automata}
\author{Henryk Fuk\'s 
      \oneaddress{
         Department of Mathematics\\
            Brock University\\
         St. Catharines, Ontario  L2S 3A1, Canada\\
         \email{hfuks@brocku.ca}
       }
   }
\date{}
\Abstract{
We present a method for construction of approximate orbits of measures
under the action of cellular automata which is complementary to the local structure
theory. The local structure theory is based on the idea of Bayesian
extension, that is, construction of a probability measure
consistent with given block probabilities and maximizing entropy.
If instead of maximizing entropy one minimizes it, one can
develop another method for construction of approximate orbits, at the heart
of which is the iteration of
finitely-dimensional maps, called minimal entropy maps.
We present numerical evidence that minimal entropy approximation
sometimes spectacularly outperforms the local structure
theory in characterizing properties of cellular automata.
Density response curve for elementary CA rule 26
is used to illustrate this claim.
}

\maketitle

\section{Introduction}
Let ${\mathcal{A}}=\{0,1,\ldots,N-1\}$ be called an \emph{alphabet}, or a \emph{symbol set}, 
and let $X={\mathcal{A}}^\mathbb{Z}$.
%The Cantor metric on $X$ is defined as $d(\mathbf{x},\mathbf{y})=2^{-k}$, where $k=\mathrm{min} \{ |i|: \mathbf{x}_i
%\neq  \mathbf{y}_i\}$. $X$ with the metric $d$ is a Cantor space, that is, compact, totally
%disconnected and perfect metric space. 
Elements of $X$ will be called configurations.
A finite sequence of elements of ${\mathcal{A}}$, $\mathbf{b}=b_1b_2\ldots, b_{n}$ will be called a \emph{block}  (or \emph{word})
 of length $n$.
Set of all blocks of elements of ${\mathcal{A}}$ of all possible lengths will be denoted by ${\mathcal{A}}^{\star}$.
\emph{Cylinder set} generated by the block $\mathbf{b}=b_1b_2\ldots b_{n}$ and anchored at $i$  is defined as
\begin{equation}
[\mathbf{b}]_i=\{ \mathbf{x}\in {\mathcal{A}}^\mathbb{Z}: \mathbf{x}_{[i,i+n)}=\mathbf{b} \}.
%\{ \mathbf{x}\in {\mathcal{A}}^\mathbb{Z}:
%x_i=b_1, x_{i+1}=b_2, \ldots, x_{i+n-1}=b_n\},
\end{equation}

The appropriate mathematical description of a distribution of configurations
is a probability measure on $X$. Cellular automata (CA) are often considered as maps in the space of such
 probability measures \cite{KurkaMaas2000,Kurka2005,Pivato2009,FormentiKurka2009}.

In this paper, we will be
interested in shift-invariant probability measures over $X$, or more precisely, 
in shift-invariant probability measures on the $\sigma$-algebra generated by elementary cylinder 
sets of $X$. Set of such measures will be denoted by $\mathfrak{M}_{\sigma}(X)$.
Detailed construction of measures in $\mathfrak{M}_{\sigma}(X)$ is described in the review
article \cite{paper50}, and interested reader is advised to consult this reference.
Nevertheless, it is not necessary to be familiar with the details of the construction
in order to follow the present paper.

The most important feature of any  measure  $\mu \in \mathfrak{M}_{\sigma}(X)$ is that it is
fully determined by measures of cylinder sets $\mu ([\mathbf{a}]_i)$ for all 
$\mathbf{a}\in {\mathcal{A}}^{\star}$, which 
we will denote by
\begin{equation}
 P(\mathbf{a})=\mu ([\mathbf{a}]_i).
\end{equation}
Note that $P(\mathbf{a})$, which we will call \emph{block probability}, is independent 
of $i$ due to shift-invariance of the measure $\mu$. Block probability $P(\mathbf{a})$ can be 
intuitively understood as the probability of occurrence of a given block $\mathbf{a}$
in the distribution of configurations.

The following theorem formally states the connection between
block probabilities and measures. It is a direct consequence of Hahn-Kolmogorov 
extension theorem. For proof the reader can consult \cite{paper50} and references therein.
\begin{thm}\label{extensionfromblocks}
 Let $P: {\mathcal{A}}^{\star} \to [0,1]$  satisfy
the conditions
\begin{align} 
P(\mathbf{b})&= \sum_{a \in {\mathcal{A}}} P(\mathbf{b}a) =\sum_{a \in \cal{G}} P(a\mathbf{b})
\,\,\,\,\,\,\,\forall {\mathbf{b} \in {\mathcal{A}}^{\star}} ,\label{cons1}\\
1&=\sum_{a \in \cal{G}} P(a).\label{cons2}
\end{align}
Then $P$ uniquely determines shift-invariant probability measure on the $\sigma$-algebra
generated by elementary cylinder sets of $X$. 
\end{thm}
The conditions (\ref{cons1}) and (\ref{cons2}) are known in the literature
as \emph{consistency conditions}.

Although the set of all block probabilities $\{P(\mathbf{a}): \mathbf{a} \in \mathcal{A}^{\star}\}$
is countable, it is still infinite, and in many practical problems, such as computer
simulations, it is often possible to keep track of only a finite number of
block probabilities. This brings an important question: if we know probabilities
of all blocks of a given length, can we reconstruct the entire measure 
\emph{approximately}? One answer to this question is well known and called
``Bayesian extension'', originally introduced 
in the context of lattice gases
\cite{Brascamp71,Fannes84}. The approximate measure produced by the Bayesian extension
is known as a ``finite-block measure'' or as ``Markov process with memory''. The
aforementioned review paper \cite{paper50} discusses details of the Bayesian
extension. The main feature of this extension is that, given a finite
set of probabilities, it constructs all other block probabilities, satisfying
consistency conditions, such that the resulting measure has the maximal entropy.

The Bayesian extension proved to be a very useful device in statistical
physics as well as in the theory of cellular automata. In 1987,  H. A. Gutowitz, J. D. Victor, 
and B. W. Knight \cite{gutowitz87a} proposed a generalization
of the mean-field theory for cellular automata based on the idea of Bayesian extension.
They called it \emph{local structure theory}. The local structure theory, recently formalized and
extended \cite{paper50}, turned out to be a very powerful tool  
for characterization  of cellular automata.

Given the success of the local structure theory, which is based on the maximal entropy
approximation, it seems quite natural to ask how useful the complementary 
approximation would be, namely the one which \emph{minimizes} the entropy instead
of maximizing it? To the knowledge of the author, no one has ever pursued
this idea, and this article is intended to partially fill this gap.

In what follows, we will investigate the minimal entropy approximation
in a configuration space over a binary alphabet, that is, assuming $\mathcal{A}=\{0,1\}$.
Although ideas presented below can be easily carried over to alphabets of higher
cardinality, the binary case is the simplest and the most elegant one,
and that is the only reason why we restrict our attention to $\mathcal{A}=\{0,1\}$.
\section{Minimal entropy extension}
Before we proceed, 
let us define $\mathbf{P}^{(k)}$ to be the column vector of all 
probabilities of blocks of length $k$ arranged in lexical order. For
example, for ${\mathcal{A}}=\{0,1\}$, the first three vectors $\mathbf{P}^{(k)}$ are
\begin{align*}
 \mathbf{P}^{(1)}&=[P(0), P(1)]^T,\\
\mathbf{P}^{(2)}&=[P(00),P(01),P(10),P(11)]^T,\\
\mathbf{P}^{(3)}&=[P(000),P(001),P(010),P(011),P(100),P(101),P(110),P(111)]^T,\\
&\cdots .
\end{align*}
Entropy of $\mathbf{P}^{(k)}$ will be defined as
\begin{equation}
h\big(\mathbf{P}^{(k)}\big)=-\sum_{\mathbf{b}\in
 {\mathcal{A}}^{k}} \widehat{P}(\mathbf{b}) \log \widehat{P}(\mathbf{b}).
\end{equation}
Suppose that for a given probability measure we know all block probabilities $\mathbf{P}^{(1)},\mathbf{P}^{(2)},\ldots,\mathbf{P}^{(n)}$. 
We want to construct block probabilities $\mathbf{P}^{(n+1)}$ 
which minimize entropy $h\big(\mathbf{P}^{(n+1)}\big)$ and are consistent with block probabilities 
$\mathbf{P}^{(1)},\mathbf{P}^{(2)},\ldots,\mathbf{P}^{(n)}$.

In order to do this, we fist must remark that not all block probabilities
which are elements of  vectors $\mathbf{P}^{(1)},\mathbf{P}^{(2)},\ldots,\mathbf{P}^{(n)}$
are independent, due to consistency conditions. In \cite{paper50}, we demonstrated
that for $\mathcal{A}=\{0,1\}$, only $2^{n-1}$ block probabilities are independent.
Which ones are declared to be independent, and which ones are treated as dependent, is
to some extent arbitrary. One choice of independent probabilities is
called \emph{short form representation} \cite{paper50}. For the binary alphabet,
in the short form representation block probabilities which have the form
$P(0\mathbf{a}0)$ are declared to be independent, and the remaining ones are treated as dependent.
For example, among elements of $\mathbf{P}^{(1)},\mathbf{P}^{(2)},\mathbf{P}^{(3)}$,
the independent probabilities are $P(0), P(00), P(000), P(010)$. The remaining ones can be expressed as
\begin{align} \label{shortform3}
%%%%%%%%%%%%%%%%%%%%%%%%%%
\left[ \begin {array}{c} 
P(001)\\
P(011)\\  
P(100) \\
P(101)\\
P(110)\\
P(111)
\end {array} \right] &= 
\left[ \begin {array}{c} 
P(00)-P(000) \\ 
P(0)-P(00) -P(010) \\  
P(00)-P(000) \\  
P(0)-2 P(00)+P(000) \\ 
P(0)-P(00)-P(010)\\  
 1-3 P(0) +2 P(00) +P(010) 
\end {array} \right]. \nonumber \\
%%%%%%%%%%%%%%%%%%%%%%%%
%%%%%%%%%%%%%%%%%%%%%%%%%
\left[ \begin {array}{c} 
P(01)\\
P(10)\\  
P(11)
\end {array} \right] &= 
\left[ \begin {array}{c} 
P(0) -P(00) \\
P(0) -P(00) \\ 
 1-2 P(0)+P(00)
\end {array} \right], \nonumber  	\\
%%%%%%%%%%%%%%%%%%%%%%%
%%%%%%%%%%%%%%%%%%%%%%%%%
 P(1) &= 1-P(0). 
%%%%%%%%%%%%%%%%%%%%%%%%
\end{align}

Coming back to our problem, if we want to construct $\mathbf{P}^{(n+1)}$
given $\mathbf{P}^{(1)},\mathbf{P}^{(2)},\ldots,\mathbf{P}^{(n)}$,
we are free to choose only the values of elements of $\mathbf{P}^{(n+1)}$
which are of the form $P(0\mathbf{a}0)$, where $\mathbf{a}\in {\mathcal{A}}^{n-1}$.
These probabilities will be denoted by $x_\mathbf{a}$, and the remaining ones 
can be expressed in terms of $x_\mathbf{a}$ and probabilities of shorter blocks,
\begin{align}\label{xdef}
 P(0\mathbf{a}0)&=x_{\mathbf{a}} ,\nonumber \\
 P(0\mathbf{a}1)&=P(0\mathbf{a})-x_{\mathbf{a}} , \nonumber \\
 P(1\mathbf{a}0)&=P(\mathbf{a}0)-P(0\mathbf{a}0)=P(\mathbf{a}0)-x_{\mathbf{a}}, \nonumber\\
 P(1\mathbf{a}1)&=P(\mathbf{a}1)-P(0\mathbf{a}1)=P(\mathbf{a}1)-(P(0\mathbf{a})-x_{\mathbf{a}}).
\end{align}
The problem is now as follows: how to choose parameters $x_{\mathbf{a}}$ in order to minimize 
entropy $h\big(\mathbf{P}^{(n+1)}\big)$?
The following theorem provides the answer.
\begin{thm}\label{mainthm}
 Suppose that $\mu$ is a shift-invariant probability measure, and $P(\mathbf{b})=\mu([\mathbf{b}]_i)$. 
Let  
% \begin{align}
%  \widehat{P}(0\mathbf{a}0)&=\widehat{x}_{\mathbf{a}} P(0\mathbf{a}),\\
%  \widehat{P}(0\mathbf{a}1)&=(1-\widehat{x}_{\mathbf{a}}) P(0\mathbf{a}),\\
%  \widehat{P}(1\mathbf{a}0)&=P(\mathbf{a}0)-\widehat{x}_{\mathbf{a}} P(0\mathbf{a}),\\
%  \widehat{P}(1\mathbf{a}1)&=P(\mathbf{a}1)-(1-\widehat{x}_{\mathbf{a}})P(0\mathbf{a}),
% \end{align}
\begin{align} \label{maintheoremformolae}
 \widehat{P}(0\mathbf{a}0)&=\widehat{x}_{\mathbf{a}} , \nonumber \\
 \widehat{P}(0\mathbf{a}1)&=P(0\mathbf{a})-\widehat{x}_{\mathbf{a}}, \nonumber  \\
 \widehat{P}(1\mathbf{a}0)&=P(\mathbf{a}0)-\widehat{x}_{\mathbf{a}},\nonumber  \\
 \widehat{P}(1\mathbf{a}1)&=P(\mathbf{a}1)-\left(P(0\mathbf{a})-\widehat{x}_{\mathbf{a}}\right),
\end{align}
where 
\begin{equation}\label{xadef}
 \widehat{x}_{\mathbf{a}}=\begin{cases} \max \big\{0,P(0\mathbf{a})-P(\mathbf{a}1)\big\}& 
\text{if $|P(\mathbf{a}1)-P(0\mathbf{a})|<|P(\mathbf{a}0)-P(0\mathbf{a})|$},\\
% \text{if $\max\left(\frac{P(0\mathbf{a})}{2},\min(P(0\mathbf{a}),P(\mathbf{a}0))\right)<\frac{P(\mathbf{a}1)+P(\mathbf{a}0)}{2}<\max(P(0\mathbf{a}),P(\mathbf{a}0))$},\\
                \min \big\{P(0\mathbf{a}),P(\mathbf{a}0)\big\} & \text{otherwise}.
 \end{cases}
\end{equation}
%
% \begin{equation}\label{xadef}
%  \widehat{x}_{\mathbf{a}}=\begin{cases} \max(0,1-\frac{P(\mathbf{a}1)}{P(0\mathbf{a})})& 
% \text{if $|P(\mathbf{a}1)-P(0\mathbf{a})|<|P(\mathbf{a}0)-P(0\mathbf{a})|$},\\
% % \text{if $\max\left(\frac{P(0\mathbf{a})}{2},\min(P(0\mathbf{a}),P(\mathbf{a}0))\right)<\frac{P(\mathbf{a}1)+P(\mathbf{a}0)}{2}<\max(P(0\mathbf{a}),P(\mathbf{a}0))$},\\
%                 \min(1,\frac{P(\mathbf{a}0)}{P(0\mathbf{a})}) & \text{otherwise}.
%  \end{cases}
% \end{equation}
Then 
\begin{equation} \label{ineqH}
 -\sum_{\mathbf{b}\in {\mathcal{A}}^{n+1}} \widehat{P}(\mathbf{b}) \log \widehat{P}(\mathbf{b})\leq  
-\sum_{\mathbf{b}\in {\mathcal{A}}^{n+1}} P(\mathbf{b}) \log P(\mathbf{b}).
\end{equation}
\end{thm}
\emph{Proof.} 
Let us first notice that $P(0\mathbf{a})$, $P(\mathbf{a}0)$, and $P(\mathbf{a}1)$ are not independent.  Consistency conditions  imply that
\begin{equation}
 P(0\mathbf{a}) \leq P(\mathbf{a}1)+P(\mathbf{a}0)\leq 1,
\end{equation}
and from there we obtain
\begin{equation}
% \delta_{\mathbf{a}} \geq \alpha_{\mathbf{a}}-\beta_{\mathbf{a}} \text{\,\,\,and\,\,\,\,}  \delta_{\mathbf{a}} \leq 1-\beta_{\mathbf{a}},
% \max( 0, \alpha_{\mathbf{a}}-\beta_{\mathbf{a}}) \leq \delta_{\mathbf{a}} \leq 1-\beta_{\mathbf{a}}.
P(0\mathbf{a})-P(\mathbf{a}0) \leq P(\mathbf{a}1) \leq 1-P(\mathbf{a}0).
\end{equation}
Denoting $\alpha_{\mathbf{a}}=P(0\mathbf{a})$, $\beta_{\mathbf{a}}=P(\mathbf{a}0)$, $\delta_{\mathbf{a}}=P(\mathbf{a}1)$, this can be written as
\begin{equation}\label{deltacond}
\alpha_{\mathbf{a}}-\beta_{\mathbf{a}} \leq \delta_{\mathbf{a}} \leq 1-\beta_{\mathbf{a}}.
\end{equation}
The right hand side of inequality (\ref{ineqH}) can be written as
\begin{gather}
  -\sum_{\mathbf{b}\in {\mathcal{A}}^{n+1}} {P}(\mathbf{b}) \log {P}(\mathbf{b})=\\
-\sum_{\mathbf{a}\in {\mathcal{A}}^{n-1}} 
 {P}(0\mathbf{a}0) \log {P}(0\mathbf{a}0)
+{P}(0\mathbf{a}1) \log {P}(0\mathbf{a}1)
+{P}(1\mathbf{a}0) \log {P}(1\mathbf{a}0)
+{P}(1\mathbf{a}1) \log {P}(1\mathbf{a}1).\nonumber
\end{gather}
Using eqs. (\ref{xdef}) this becomes
\begin{gather}
  -\sum_{\mathbf{b}\in {\mathcal{A}}^{n+1}} {P}(\mathbf{b}) \log {P}(\mathbf{b})=
\sum_{\mathbf{a}\in {\mathcal{A}}^{n-1}} H_a(x_{\mathbf{a}}),
\end{gather}
where we define 
\begin{align}
 H_a(x_{\mathbf{a}})=&
-x_{\mathbf{a}}  \log x_{\mathbf{a}} 
-(\alpha_{\mathbf{a}}-x_{\mathbf{a}})  \log \left(\alpha_{\mathbf{a}}-x_{\mathbf{a}}\right)\nonumber \\
-&(\beta_{\mathbf{a}}-x_{\mathbf{a}}) \log (\beta_{\mathbf{a}}-x_{\mathbf{a}})
-\left(\delta_{\mathbf{a}}-(\alpha_{\mathbf{a}}-x_{\mathbf{a}}) \right)\log \left(\delta_{\mathbf{a}}-(\alpha_{\mathbf{a}}-x_{\mathbf{a}})\right).\
\end{align}
Function $H_a(x_{\mathbf{a}})$ is concave, and $x_{\mathbf{a}}$ can only take values from some closed interval $[x_{\mathbf{a},1},x_{\mathbf{a},2}]$. For this
reason, $H_a(x_{\mathbf{a}})$ reaches minimum at one of the endpoints of the interval. We will show that the minimum occurs 
precisely at $x_{\mathbf{a}}=\widehat{x}_{\mathbf{a}}$, where $\widehat{x}_{\mathbf{a}}$ is defined in eq. (\ref{xadef}).
First, let us determine the values of the endpoints $x_{\mathbf{a},1},x_{\mathbf{a},2}$. In order to do this, 
note that obviously
$x_{\mathbf{a}} \in [0,1]$. By consistency conditions, 
\begin{align}
P(\mathbf{a}1)&=P(0\mathbf{a}1)+P(1\mathbf{a}1),\\
P(\mathbf{a}0)&=P(0\mathbf{a}0)+ P(1\mathbf{a}0),\\
P(0\mathbf{a})&=P(0\mathbf{a}0)+ P(0\mathbf{a}1),
\end{align}
and therefore
\begin{align}
P(\mathbf{a}1)& \geq P(0\mathbf{a}1),\\
P(\mathbf{a}0)& \geq P(0\mathbf{a}0),\\
P(0\mathbf{a})& \geq P(0\mathbf{a}0).
\end{align}
Using eqs. (\ref{xdef}) and the notation $\alpha_{\mathbf{a}}=P(0\mathbf{a})$, 
$\beta_{\mathbf{a}}=P(\mathbf{a}0)$, $\delta_{\mathbf{a}}=P(\mathbf{a}1)$, this becomes
\begin{align}
\delta_{\mathbf{a}}& \geq \alpha_{\mathbf{a}}-x_{\mathbf{a}},\\
\beta_{\mathbf{a}}& \geq x_{\mathbf{a}} ,\\
\alpha_{\mathbf{a}}& \geq x_{\mathbf{a}}.
\end{align}
Solving the above system of inequalities for $x_{\mathbf{a}}$ we obtain
\begin{equation}
 \alpha_{\mathbf{a}}-\delta_{\mathbf{a}}      \leq x_{\mathbf{a}} \leq \min \left\{ \alpha_{\mathbf{a}}, \beta_{\mathbf{a}}\right\}.
\end{equation}
Since $x_{\mathbf{a}} \in [0,1]$, we obtain the following expression for the endpoints of the interval $[x_{\mathbf{a},1},x_{\mathbf{a},2}]$,
\begin{equation}\label{xa12}
 x_{\mathbf{a},1}=\max\left\{0,\alpha_{\mathbf{a}}-\delta_{\mathbf{a}}\right\}, \,\,\,\,\,\,\, x_{\mathbf{a},2}=\min\left\{\alpha_{\mathbf{a}},\beta_{\mathbf{a}} \right\}.
\end{equation}

Suppose now that we fix both $\mathbf{a}$ and $\alpha_{\mathbf{a}}$.
Let us consider separately the four cases described in the table below.
\begin{center}
\begin{tabular}{l|l|l}
 & $\beta_{\mathbf{a}} < \alpha_{\mathbf{a}}$ & $\beta_{\mathbf{a}} \geq \alpha_{\mathbf{a}}$\\ \hline
$\delta_{\mathbf{a}} < \alpha_{\mathbf{a}}$ & $x_{\mathbf{a},1}=\alpha_{\mathbf{a}}-\delta_{\mathbf{a}}$ & $x_{\mathbf{a},1}=\alpha_{\mathbf{a}}-\delta_{\mathbf{a}}$\\
                  &  $x_{\mathbf{a},2}=\beta_{\mathbf{a}}$   & $x_{\mathbf{a},2}=\alpha_{\mathbf{a}}$\\ \hline
$\delta_{\mathbf{a}} \geq \alpha_{\mathbf{a}}$ & $x_{\mathbf{a},1}=0$ & $x_{\mathbf{a},1}=0$ \\
                     & $x_{\mathbf{a},2}=\beta_{\mathbf{a}}$ & $x_{\mathbf{a},2}=\alpha_{\mathbf{a}}$
\end{tabular}
\end{center}
We will  determine the sign of $H_{\mathbf{a}}(x_{\mathbf{a},1})-H_{\mathbf{a}}(x_{\mathbf{a},2})$. If $H_{\mathbf{a}}(x_{\mathbf{a},1})-H_{\mathbf{a}}(x_{\mathbf{a},2})<0$, then the minimum
of $H_a$ occurs at  $x_{\mathbf{a},1}$, otherwise at $x_{\mathbf{a},2}$. To avoid notational clutter, 
we will drop the index $\mathbf{a}$ from $\alpha_{\mathbf{a}}$, $\beta_{\mathbf{a}}$, $\delta_{\mathbf{a}}$.

\textbf{Case 1:} $\beta < \alpha$, $\delta < \alpha$.\\
We have
\begin{equation}
H_{\mathbf{a}}(x_{\mathbf{a},1})-H_{\mathbf{a}}(x_{\mathbf{a},2})=\beta\log \beta + (\alpha-\beta)\log (\alpha - \beta)
-(\alpha-\delta) \log(\alpha-\delta) - \delta \log \delta.
\end{equation}
Defining $f_p(x)=x \log x + (p-x)\log(p-x)$ we can write
\begin{equation}
H_{\mathbf{a}}(x_{\mathbf{a},1})-H_{\mathbf{a}}(x_{\mathbf{a},2})=f_\alpha(\beta)-f_\alpha(\delta).
 \end{equation}
The function $f_\alpha(x)$, defined on interval $x\in (0, \alpha)$, reaches minimum at $x=\alpha/2$, and has the property
$f_\alpha(x)=f_\alpha(\alpha-x)$. This means that $f_\alpha(\delta)>f_\alpha(\beta)$, and thus $H_{\mathbf{a}}(x_{\mathbf{a},1})-H_{\mathbf{a}}(x_{\mathbf{a},2})<0$,  if and only if 
\begin{equation} \label{case1ineq}
 \left|\delta-\frac{\alpha}{2}\right| > \left|\beta-\frac{\alpha}{2}\right|.
\end{equation}

\textbf{Case 2:} $\beta < \alpha$, $\delta \geq \alpha$.\\
We have
\begin{align}
H_{\mathbf{a}}(x_{\mathbf{a},1})-H_{\mathbf{a}}(x_{\mathbf{a},2})&=-\alpha \log \alpha
-(\delta-\alpha) \log (\delta-\alpha)\\
+&(\alpha-\beta) \log(\alpha-\beta)
+(\beta-\alpha+\delta)\log(\beta-\alpha+\delta)= \nonumber\\
&f_\delta(\alpha-\beta)-f_\delta(\alpha) \nonumber.
\end{align}
For the same reason as before, $H_{\mathbf{a}}(x_{\mathbf{a},1})-H_{\mathbf{a}}(x_{\mathbf{a},2})<0$  if and only if
\begin{equation}\label{case2ineq}
 \left|\alpha-\frac{\delta}{2}\right| > \left|\alpha-\beta-\frac{\delta}{2}\right|.
\end{equation}
\textbf{Case 3:} $\beta \geq \alpha$, $\delta < \alpha$.\\
We have
\begin{align}
H_{\mathbf{a}}(x_{\mathbf{a},1})-H_{\mathbf{a}}(x_{\mathbf{a},2})&=
-(\alpha-\delta) \log(\alpha-\delta)\\
-&(\beta-\alpha+\delta)\log(\beta-\alpha+\delta)
+\alpha \log \alpha
+(\beta-\alpha) \log (\beta-\alpha) \nonumber \\
=&f_\beta(\alpha)-f_\beta(\alpha-\delta) \nonumber
\end{align}
Again, by the property of $f_\beta$ discussed under Case 1, $H_{\mathbf{a}}(x_{\mathbf{a},1})-H_{\mathbf{a}}(x_{\mathbf{a},2})<0$
 is equivalent to
\begin{equation}\label{case3ineq}
 \left|\alpha-\delta-\frac{\beta}{2}\right| > \left|\alpha-\frac{\beta}{2}\right|.
\end{equation}
\textbf{Case 4:} $\beta \geq \alpha$, $\delta \geq \alpha$.\\
We have
\begin{align}
H_{\mathbf{a}}(x_{\mathbf{a},1})-H_{\mathbf{a}}(x_{\mathbf{a},2})&=
-\beta \log \beta
-(\delta-\alpha) \log (\delta-\alpha)  \\
+&(\beta-\alpha) \log (\beta-\alpha)
+\delta \log \delta \nonumber\\
=&g(\delta)-g(\beta), \nonumber
\end{align}
where $g(y)=y \log y - (y-\alpha) \log (y-\alpha)$. Since for $y \in (0,\alpha)$
\begin{equation}
 g^{\prime}(y)=\log y - \log (y-\alpha)>0,
\end{equation}
$h(y)$ is increasing in $(0,\alpha)$. This means that $g(\delta)>g(\beta)$, or equivalently
 $H_{\mathbf{a}}(x_{\mathbf{a},1})-H_{\mathbf{a}}(x_{\mathbf{a},2})<0$,
 is satisfied if and only if
\begin{equation}\label{case4ineq}
\delta<\beta. 
\end{equation}
\begin{figure}
 \begin{center}
     \includegraphics[scale=1.1]{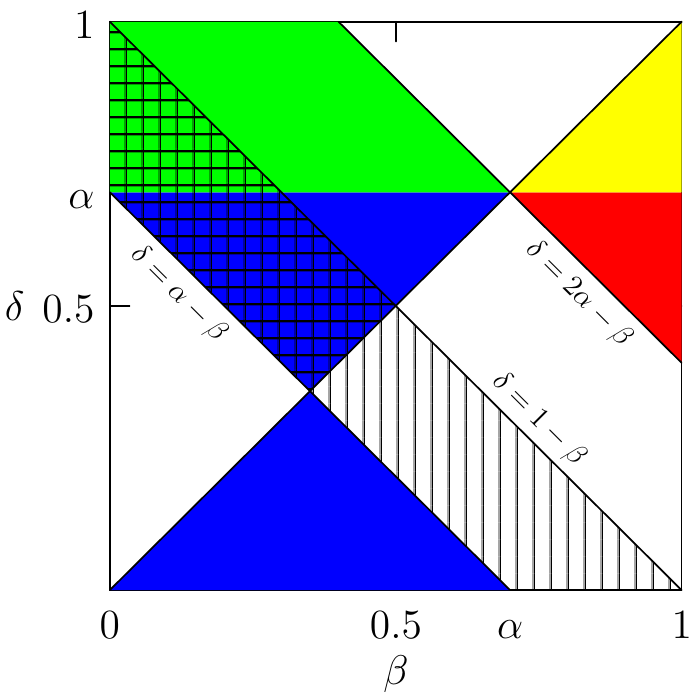}\includegraphics[scale=1.1]{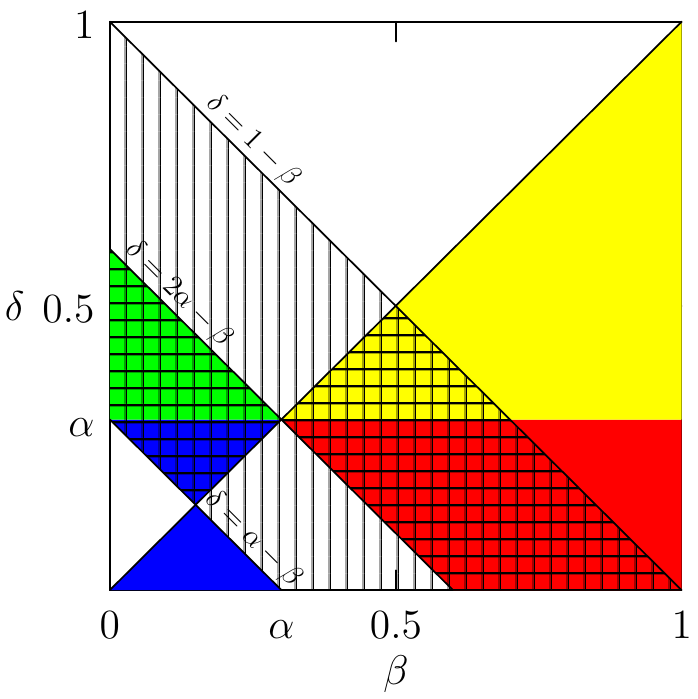}
   \end{center}
\caption{Solutions of inequalities (\ref{case1ineq}),  (\ref{case2ineq}), (\ref{case3ineq}) 
and (\ref{case4ineq})  in $\beta$-$\delta$ space, shown, respectively, in blue, green, red, and yellow color.
Region with vertical hatching represents solution of inequality (\ref{deltacond}), and the region with
horizontal hatching represents parameters for which the minimum of $H_a$ occurs at $x_{\mathbf{a},1}$.
Two scenarios of are shown, corresponding to $a>0.5$ (left) and  $a<0.5$ (right).
}\label{fig1}
\end{figure}
\begin{figure}
 \begin{center}
     \includegraphics[scale=1.1]{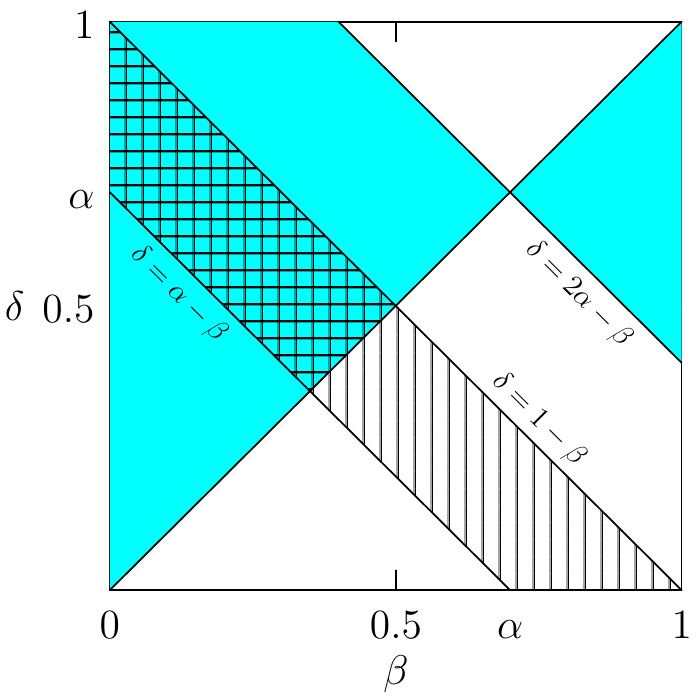}\includegraphics[scale=1.1]{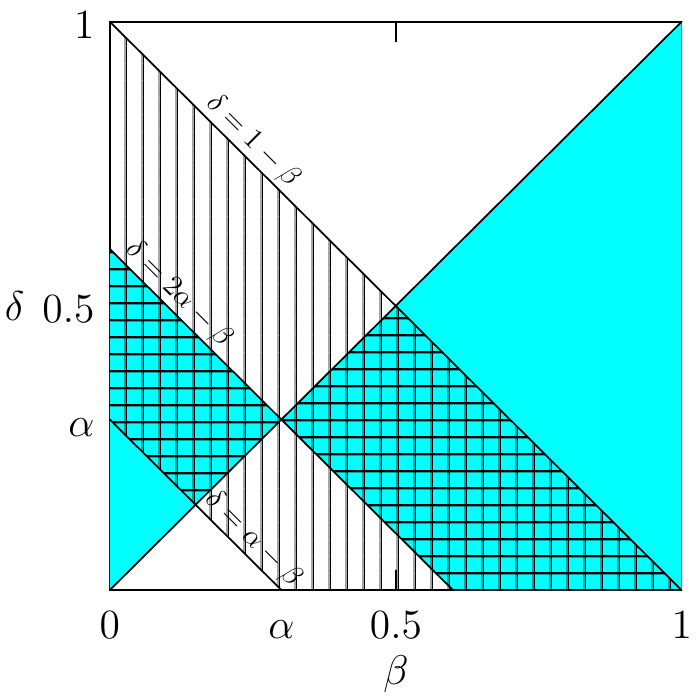}
   \end{center}
\caption{Simplification of Figure 1. Solutions of inequalities (\ref{case1ineq}),  (\ref{case2ineq}), (\ref{case3ineq}) 
and (\ref{case4ineq}) have been replaced by solution of $|\delta-\alpha|<|\beta-\alpha|$, shown in cyan color.}\label{fig2}
\end{figure}
%%%%%%

We obtained four inequalities (\ref{case1ineq}),  (\ref{case2ineq}), (\ref{case3ineq}), and (\ref{case4ineq}) for four cases. 
We plotted in Figure~\ref{fig1}  solutions of these four inequalities  in $\beta$-$\delta$ space
using different colors for each case. One can, however, combine all four cases and describe them by one simple inequality
taking into account the fact that only values of $(\beta, \delta)$ marked by vertical hashing are possible, due to condition (\ref{deltacond}).
This simple inequality combining all four cases (subject to condition (\ref{deltacond})) is
\begin{equation}
|\delta-\alpha|<|\beta-\alpha|,
\end{equation}
as one can easily verify graphically by comparing Figures \ref{fig1} and \ref{fig2}.

To summarize our findings, we demonstrated that the minimum of $H_{\mathbf{a}}$ occurrs at $\widehat{x}_{\mathbf{a}}$, where
\begin{equation}
 \widehat{x}_{\mathbf{a}}=\begin{cases} x_{\mathbf{a},1},& 
\text{if $|\delta_{\mathbf{a}}-\alpha_{\mathbf{a}}|<|\beta_{\mathbf{a}}-\alpha_{\mathbf{a}}|$},\\
                x_{\mathbf{a},2} & \text{otherwise},
 \end{cases}
\end{equation}
and where $x_{\mathbf{a},1}$ and $x_{\mathbf{a},2}$ are defined in eq. (\ref{xa12}).
This is precisely eq. (\ref{xadef}), and Theorem \ref{mainthm} then follows directly. $\square$

\section{Minimal entropy approximation for measures}
Using Theorem \ref{mainthm}, we can now construct approximation of a probability measure
complementary to the Bayesian approximation. Let us define
\begin{equation}
\Upsilon(\alpha,\beta,\delta)=\begin{cases} \max \big\{0,\alpha-\delta\big\}& 
\text{if $|\delta-\alpha|<|\beta-\alpha|$},\\
% \text{if $\max\left(\frac{P(0\mathbf{a})}{2},\min(P(0\mathbf{a}),P(\mathbf{a}0))\right)<\frac{P(\mathbf{a}1)+P(\mathbf{a}0)}{2}<\max(P(0\mathbf{a}),P(\mathbf{a}0))$},\\
                \min \big\{\alpha,\beta\big\} & \text{otherwise}.
 \end{cases}
\end{equation}
and
\begin{align}
\Upsilon_{0,0}(\alpha,\beta,\delta)&=\Upsilon(\alpha,\beta,\delta),\\
\Upsilon_{0,1}(\alpha,\beta,\delta)&=\alpha-\Upsilon(\alpha,\beta,\delta),\\
\Upsilon_{1,0}(\alpha,\beta,\delta)&=\beta-\Upsilon(\alpha,\beta,\delta),\\
\Upsilon_{1,1}(\alpha,\beta,\delta)&=\delta-\alpha+\Upsilon(\alpha,\beta,\delta).
\end{align}
Using this notation and eq. (\ref{maintheoremformolae}), one can now express probabilities of $(k+1)$-blocks by probabilities
of $k$-blocks, by writing
\begin{equation}
P(a_1a_2 \ldots a_{k+1}) \approx \Upsilon_{a_1,a_{k+1}}\Big( P(0a_2 \ldots a_{k}),
P(a_2 \ldots a_{k}0),P(a_2 \ldots a_{k}1)\Big).
\end{equation}
The above approximation will be called \emph{minimal entropy approximation}.
We can, of course, repeat this process, and approximate $(k+2)$-block probabilities by 
$(k+1)$-block probabilities,
and, by applying the approximation again,  $(k+1)$-block probabilities by $k$-block probabilities. 
For a given integer $k$, by recursive application of the minimal entropy approximation, any block probability of length
$p>k$ can be expressed by probabilities of length $k$. 
The following proposition states this more formally.
\begin{prop}\label{minextprop}
Let $\mu \in \mathfrak{M}_{\sigma}(X)$ be a measure with associated block probabilities 
$P: {\mathcal{A}}^{\star} \to [0,1]$,
$P(\mathbf{b})=\mu([\mathbf{b}]_i)$ for all $i\in \mathbb{Z}$ and $\mathbf{b}\in {\mathcal{A}}^{\star}$.
For $k>0$,  define
$\widehat{P}: {\mathcal{A}}^{\star} \to [0,1]$ recursively so that
\begin{equation} \label{hatdef}
 \widehat{P}(a_1a_2 \ldots a_{p})=\\
\begin{cases}
P(a_1a_2 \ldots a_{p}) & \mathrm{if\,\,} p \leq k ,\\[0.5em]
\Upsilon_{a_1,a_p}\Big( \widehat{P}(0a_2 \ldots a_{p-1}),\widehat{P}(a_2 \ldots a_{p-1}0),\widehat{P}(a_2 \ldots a_{p-1}1)\Big) 
& \mathrm{if\,\,} p>k.
\end{cases}
\end{equation}
Then $\widehat{P}$ determines a shift-invariant probability measure $\widehat{\mu}^{(k)} \in \mathfrak{M}_\sigma(X)$,
 to
be called \emph{minimal entropy approximation of $\mu$ of order $k$}. 
\end{prop}
The proof that $\widehat{P}$ determines a measure is a direct consequence of the definition of $\widehat{P}$.

It is intuitively clear that as the order of the minimal entropy approximations increases, 
the quality of the approximation should
increase too. The following proposition formalizes this observation.
 \begin{prop}
  The sequence of $k$-th order minimal entropy approximations of
$\mu \in \mathfrak{M}_\sigma(X)$ weakly converges to  $\mu$ as $k \to \infty$.
 \end{prop}
\emph{Proof:}
Let $n>0$, $\mathbf{b} \in {\mathcal{A}}^n$ and let
$\widehat{P}_k(\mathbf{b})=\widehat{\mu}^{(k)}([\mathbf{b}]_0)$, 
 $P(\mathbf{b})=\mu([\mathbf{b}]_0)$.
Since for $k \geq n$ $\widehat{P}_k(\mathbf{b})=P(\mathbf{b})$,
we obviously have $\lim_{k \to \infty} \widehat{P}_k(\mathbf{b})=P(\mathbf{b})$.
Since cylinder sets constitute convergence determining class for measures in $\mathfrak{M}_\sigma(X)$,
convergence of block probabilities is equivalent to weak convergence. 
This leads to the conclusion that
$\widetilde{\mu}^{(k)} \Rightarrow \mu$.  $\square$

Measures $\mu$ for which $\widetilde{\mu}^{(k)} = \mu$ will be called \emph{$k$-th order measures of 
minimal entropy}. Set of such measures over $X$ will be denoted by $\mathfrak{M}^{(k)}_{\mathrm{ME}}(X)$.
Obviously these measures are shift-invariant, $\mathfrak{M}^{(k)}_{\mathrm{ME}}(X) \subset \mathfrak{M}^{(k)}_{\sigma}(X)$.

\section{Orbits of measures under the action of cellular automata}

Let $w: \mathcal{A} \times \mathcal{A}^{2r+1} \to [0,1]$, whose values are denoted by $w(a|\mathbf{b})$
for $a \in \mathcal{A}$, $\mathbf{b} \in \mathcal{A}^{2r+1}$, satisfying
$\sum_{a \in \mathcal{A}} w(a|\mathbf{b})=1$, be called \emph{local transition function}
of \emph{radius} $r$, and its values will be called \emph{local transition probabilities}.
\emph{Probabilistic cellular automaton}  with local 
transition function $w$ is a map $F: \mathfrak{M}_{\sigma}(X) \to \mathfrak{M}_{\sigma}(X)$ defined as
\begin{equation} \label{rulefed}
(F\mu)([\mathbf{b}]_i)=\sum_{\mathbf{a}\in \mathcal{A}^{|\mathbf{b}|+2r}} w(\mathbf{a}| \mathbf{b}) \mu([\mathbf{a}]_{i-r})
\mathrm{\,\, for\,\, all\,\,}  i \in \mathbb{Z}, \mathbf{b} \in \mathcal{A}^{\star},
\end{equation}
where we define
\begin{equation} \label{defw}
 w(\mathbf{a}| \mathbf{b}) = \prod_{j=1}^{|\mathbf{a}|} w(a_j|b_jb_{j+1}\ldots b_{j+2r}).
\end{equation}
When the function $w$ takes values in the set $\{0,1\}$, the corresponding cellular automaton is called 
\emph{deterministic CA}. 

For any probabilistic measure $\mu \in \mathfrak{M}_{\sigma}(X)$, we define the orbit of $\mu$ under $F$ as
\begin{equation}
 \{  F^n \mu \}_{n=0}^{\infty}.
\end{equation}
Excluding trivial cases, computing the orbit of a measure under a given CA is very difficult, and no
general method is known. We will, therefore, propose a method for approximating orbits
based on the minimal entropy approximation.

Let us first define the \emph{entropy minimizing operator} of order $k$, denoted by $\Psi^{(k)}$, to be a map
from $\mathfrak{M}_\sigma (X)$ to $\mathfrak{M}_{\mathrm{ME}}^{(k)} (X)$ such that
\begin{equation}
\Psi^{(k)} \mu=\widehat{\mu}^{(k)},
\end{equation}
where $\widehat{\mu}^{(k)}$ is the measure defined in Proposition~\ref{minextprop}.
Note that the operator $\Psi^{(k)}$ is indempontent, that is, $\Psi^{(k)} \Psi^{(k)} \mu=\Psi^{(k)} \mu$.
This allows us to construct approximate orbit of a measure $\mu$ under the action of $F$
by simply replacing $F$ by $\Psi^{(k)} F \Psi^{(k)}$. 
The sequence
\begin{equation} \label{approxorbit}
\left\{ \left(\Psi^{(k)} F \Psi^{(k)} \right)^n \mu \right\}_{n=0}^{\infty}
\end{equation}
will be called the \emph{minimal entropy approximation} of level $k$ of the exact orbit $\{  F^n \mu \}_{n=0}^{\infty}$.
Note that all terms of this sequence are mesures of minimal entropy, thus the entire  approximate
 orbit lies in $\mathfrak{M}_{\mathrm{ME}}^{(k)}(X)$.

Just like for the local structure approximation, the minimal entropy approximation approximates the actual orbit 
increasingly well as $k$ increases. In fact, we will prove that every point of the approximate orbit 
weakly converges to the corresponding point of the exact orbit. 

\begin{prop} 
Let $k$ be a positive integer and $\mathbf{b} \in {\mathcal{A}}^{\star}$.
If $k\geq |\mathbf{b}|+2r$, then
\begin{equation} \label{eqpsi}
F \mu ([\mathbf{b}])=F \Psi^{(k)} \mu ([\mathbf{b}])=\Psi^{(k)} F \mu ([\mathbf{b}]). 
\end{equation}
\end{prop}
\emph{Proof.} To prove it, note that $\mu([\mathbf{a}])=\widehat{\mu}^{(k)}([\mathbf{a}])$ for all blocks 
$\mathbf{a}$ of length up to $k$. The first equality of (\ref{eqpsi}) can be written as
\begin{equation}
\sum_{\mathbf{a}\in \mathcal{A}^{|\mathbf{b}|+2r}} w(\mathbf{a}| \mathbf{b}) \mu([\mathbf{a}])
=
\sum_{\mathbf{a}\in \mathcal{A}^{|\mathbf{b}|+2r}} w(\mathbf{a}| \mathbf{b}) \widehat{\mu}^{(k)}([\mathbf{a}]).
\end{equation}
The equality holds when $|\mathbf{a}|\leq k$, that is, $|\mathbf{b}|+2r \leq k$.

The second equality of (\ref{eqpsi}) is a result of the fact that the $\Psi^{(k)}$ operator only modifies probabilities of blocks of length greater than $k$.
Since $k\geq |\mathbf{b}|+2r$, we have $|\mathbf{b}|<k$ and therefore $F \mu ([\mathbf{b}])=\Psi^{(k)} F \mu ([\mathbf{b}])$.  $\square$ 

Now let us note that  $F^n$ can be viewed as a cellular automaton rule of radius $nr$, 
thus when $k\geq |\mathbf{b}|+2nr$, we have $F^n \mu ([\mathbf{b}])=F^n \Psi^{(k)} \mu ([\mathbf{b}])$. We can insert
arbitrary number of  $\Psi^{(k)}$ operators on the right hand side anywhere we want, and nothing will change, 
because $\Psi^{(k)}$ does not modify relevant block probabilities. This yields an immediate corollary.
\begin{cor}
Let $k$ and $n$ be  positive integers and $\mathbf{b} \in {\mathcal{A}}^{\star}$.
If $k\geq |\mathbf{b}|+2nr$, then
$$ F^n \mu ([\mathbf{b}])=\left( \Psi^{(k)} F \Psi^{(k)} \right)^n \mu ([\mathbf{b}]).$$
\end{cor}
This means that for a given $n$, measures of cylinder sets in the approximate 
measure $\left( \Psi^{(k)} F \Psi^{(k)} \right)^n \mu$
coincide with measures of cylinder sets in $F^n \mu$ for sufficiently large $k$. Because cylinder sets constitute 
convergence determining class for measures, we obtain
the following result.
\begin{thm} \label{lstconveergence}
 Let $F$ be a cellular automaton, $\mu \in \mathfrak{M}_{\sigma}(X)$ be a shift-invariant measure, 
and $\nu_n^{(k)}$ be a
minimal entropy approximation of order $k$ of the measure $F^n \mu$, i.e., $\nu_n^{(k)}=\left( \Psi^{(k)} F \Psi^{(k)} \right)^n \mu$.  Then 
for any positive integer $n$, 
$\nu_n^{(k)} \Rightarrow F^n \mu$ as $k \to \infty$.
\end{thm}

%Note that the the minimal entropy approximation is an idempotent operation, meaning that 
%$\widetilde{\widetilde{\mu}}=\widetilde{\mu}$. 
\section{Minimal entropy maps}
Minimal entropy measures can be  entirely described by
specifying a finite number of block probabilities. We will use this feature
to constructs a finite-dimensional map which approximates the action of a CA rule on
shift-invariant measures.
If  $\nu_n^{(k)}=\left( \Psi^{(k)} F \Psi^{(k)} \right)^n \mu$, then $\nu_n^{(k)}$ satisfies recurrence equation
\begin{equation}
  \nu_{n+1}^{(k)}= \Psi^{(k)} F \Psi^{(k)} \nu_{n}^{(k)}.
\end{equation}
On both sides of this equation we have measures  in $\mathfrak{M}_{\mathrm{ME}}^{(k)}(X)$, and these are completely determined
by probabilities of blocks of length $k$. If $|\mathbf{b}|=k$, we obtain
\begin{equation}
  \nu_{n+1}^{(k)}([\mathbf{b}])= \Psi^{(k)} F \Psi^{(k)} \nu_{n}^{(k)}([\mathbf{b}]),
\end{equation}
and, since $\Psi^{(k)}$ does not modify probabilities of blocks of length $k$, this simplifies to 
\begin{equation}
  \nu_{n+1}^{(k)}([\mathbf{b}])= F \Psi^{(k)} \nu_{n}^{(k)}([\mathbf{b}]).
\end{equation}
By the definition of $F$,
\begin{equation}
  \nu_{n+1}^{(k)}([\mathbf{b}])= 
\sum_{\mathbf{a}\in \mathcal{A}^{|\mathbf{b}|+2r}} w(\mathbf{a}| \mathbf{b})  
\left( \Psi^{(k)} \nu_{n}^{(k)} \right)
([\mathbf{a}]).
\end{equation}
To simplify the notation, let us define $Q_n(\mathbf{b})= \nu_{n}^{(k)}([\mathbf{b}])$,
and, consistent with definition in eq. (\ref{hatdef}),  $\widehat{Q}_n(\mathbf{a})=\left( \Psi^{(k)} \nu_{n}^{(k)} \right)
([\mathbf{a}])$.
 Then we can
rewrite the previous equation to take the form
\begin{equation}\label{minentcomp}
 Q_{n+1}(\mathbf{b})=\sum_{\mathbf{a}\in \mathcal{A}^{|\mathbf{b}|+2r}} w(\mathbf{a}| \mathbf{b})  
\widehat{Q}_n(\mathbf{a}).
\end{equation}
Note that by eq. (\ref{hatdef}), $\widehat{Q}_n(\mathbf{a})$ depends only on probabilities of blocks
of length $k$. If we thus arrange $Q_n(\mathbf{b})$  for all $\mathbf{b} \in \mathcal{A}^k$ in lexicographical order
  to form a vector $\mathbf{Q}_n$, we will obtain
\begin{equation} \label{minentmap}
 \mathbf{Q}_{n+1} = U^{(k)} \left(\mathbf{Q}_{n}\right),
\end{equation} 
where $U^{(k)}: [0,1]^{|\mathcal{A}|^k} \to [0,1]^{|\mathcal{A}|^k}$ has components defined by eq. (\ref{minentcomp}).
We will call this map \emph{an entropy minimizing map of order $k$}.
\section{Example: elementary CA rule 26}
As an example, consider rule 26 given by
\begin{align} \label{wfor26}
w(1|000)=0, \, w(1|001)=1, \, w(1|010)=0, \, w(1|011)=1, \nonumber \\
w(1|100)=1, \, w(1|101)=0, \, w(1|110)=0, \, w(1|111)=0,
\end{align}
and suppose we wish to construct minimal entropy map  of order 2 for this rule.
Let $P_n(\mathbf{b})=F^n \mu([\mathbf{b}])$. Using eq. (\ref{rulefed}) we obtain for $r=1$, $|b|=3$
\begin{equation}
 P_{n+1}(\mathbf{b})=\sum_{\mathbf{a}\in \mathcal{A}^{5}} 
 w(\mathbf{a}| \mathbf{b}) P_{n}(\mathbf{a}).
\end{equation}
Using definition of $w(\mathbf{a}| \mathbf{b})$ given in eq. (\ref{defw})
and transition probabilities given in eq. (\ref{wfor26}) we obtain
\begin{align} \label{r26exactk2}
 P_{n+1}(00)&=
P_n(0000)+
P_n(0101)+
P_n(1010)+
P_n(1101)+
P_n(1110)+
P_n(1111) ,\nonumber \\
P_{n+1}(01)&=
P_n(0001)+
P_n(0100)+
P_n(1011)+
P_n(1100),\nonumber \\
P_{n+1}(10)&=
P_n(0010)+
P_n(0110)+
P_n(0111)+
P_n(1000),\nonumber \\
P_{n+1}(11)&=
P_n(0011)+
P_n(1001).
\end{align}
This set of equations describes exact relationship between block probabilities at step $n+1$
and block probabilities at step $n$. Note that $3$-block probabilities at step $n+1$ are given in terms of $5$-blocks
probabilities at step $n$, thus it is not possible to iterate these equations.

Minimal entropy map of order 2 (eq.  \ref{minentmap}) can be obtained by simply replacing $P$ by $Q$
and  placing the operator $\widehat{\mbox{\,\,\,\,}}$ over probabilities on the right hand side of  eq.
(\ref{r26exactk2}). This yields
\begin{align} \label{r26minentk2}
  Q_{n+1}(00)&=
\widehat{Q}_n(0000)+
\widehat{Q}_n(0101)+
\widehat{Q}_n(1010)+
\widehat{Q}_n(1101)+
\widehat{Q}_n(1110)+
\widehat{Q}_n(1111) ,\nonumber \\
Q_{n+1}(01)&=
\widehat{Q}_n(0001)+
\widehat{Q}_n(0100)+
\widehat{Q}_n(1011)+
\widehat{Q}_n(1100),\nonumber \\
Q_{n+1}(10)&=
\widehat{Q}_n(0010)+
\widehat{Q}_n(0110)+
\widehat{Q}_n(0111)+
\widehat{Q}_n(1000),\nonumber \\
Q_{n+1}(11)&=
\widehat{Q}_n(0011)+
\widehat{Q}_n(1001).
\end{align}
Using eq. (\ref{hatdef}) with $k=2$, one can express $\widehat{Q}_n(a_1a_2a_3a_4)$ in terms of 
2-block probabilities. For example, 
\begin{multline}
\widehat{Q}_n(0000)=\Upsilon \big(\widehat{Q}_n(000),\widehat{Q}_n(000),\widehat{Q}_n(001)\big)=
\widehat{Q}_n(000)\\=\Upsilon \big(\widehat{Q}_n(00),\widehat{Q}_n(00),\widehat{Q}_n(01)\big)=\widehat{Q}_n(00)
=Q_n(00).
\end{multline}
Similarly one obtains
\begin{equation}
 \widehat{Q}_n(0101)=Q_n(01), \,\,\,\, \widehat{Q}_n(1010)=Q_n(10),\,\,\,\, \widehat{Q}_n(1111)=Q_n(11).
\end{equation}
All other $\widehat{Q}_n(a_1a_2a_3a_4)$ are equal to 0. This simplifies eq. (\ref{r26minentk2}) to
\begin{align} 
  Q_{n+1}(00)&=
Q_n(00)+
Q_n(01)+Q_n(10)+Q_n(11) ,\nonumber \\
Q_{n+1}(01)&=0,\nonumber \\
Q_{n+1}(10)&=0,\nonumber \\
Q_{n+1}(11)&=0.
\end{align}
This defines a minimal entropy map  $U^{(k)}: [0,1]^{4} \to [0,1]^{4}$ (cf. eq. \ref{minentmap})   which can be iterated, albeit in this case, it is a trivial map,
which after one iteration reaches the fixed point $(1,0,0,0)$, because $ Q_n(00)+
Q_n(01)+Q_n(10)+Q_n(11)=1$. We need higher order approximation in order to obtain a more ``interesting'' map.

When $k=3$, we follow the same procedure as for the $k=2$ case discussed above.
If we write  eq. (\ref{minentcomp}) for all possible $\mathbf{b} \in {\cal A}^3$, we will have on the left hand sides 
 eight block probabilities $Q(b_1b_2b_3)$, thus the resulting minimal entropy map will be 8-dimensional,
\begin{align}\label{rule26-k3}
Q_{n+1}(000)&=
\widehat{Q}_n(00000)+
\widehat{Q}_n(01010)+
\widehat{Q}_n(10101)+
\widehat{Q}_n(11010)+
\widehat{Q}_n(11101)\nonumber \\
 &\mbox{\hspace{4em}}+
\widehat{Q}_n(11110)+
\widehat{Q}_n(11111), \nonumber \\
Q_{n+1}(001)&=
\widehat{Q}_n(00001)+
\widehat{Q}_n(01011)+
\widehat{Q}_n(10100)+
\widehat{Q}_n(11011)+
\widehat{Q}_n(11100), \nonumber \\
Q_{n+1}(010)&=
\widehat{Q}_n(00010)+
\widehat{Q}_n(01000)+
\widehat{Q}_n(10110)+
\widehat{Q}_n(10111)+
\widehat{Q}_n(11000), \nonumber \\
Q_{n+1}(011)&=
\widehat{Q}_n(00011)+
\widehat{Q}_n(01001)+
\widehat{Q}_n(11001), \nonumber \\
Q_{n+1}(100)&=
\widehat{Q}_n(00101)+
\widehat{Q}_n(01101)+
\widehat{Q}_n(01110)+
\widehat{Q}_n(01111)+
\widehat{Q}_n(10000), \nonumber \\
Q_{n+1}(101)&=
\widehat{Q}_n(00100)+
\widehat{Q}_n(01100)+
\widehat{Q}_n(10001), \nonumber \\
Q_{n+1}(110)&=
\widehat{Q}_n(00110)+
\widehat{Q}_n(00111)+
\widehat{Q}_n(10010), \nonumber \\
Q_{n+1}(111)&=
\widehat{Q}_n(10011).
\end{align}
On the right hand side, we have 32 block probabilities which have to be
expressed in terms of 3-block probabilities by using eq. (\ref{hatdef}) with $k=3$.
Some of these will simplify to a single 3-block probability, e.g.,
\begin{equation}
\widehat{Q}_n(00000)= \Upsilon\big(\widehat{Q}_n(0000),\widehat{Q}_n(0000),\widehat{Q}_n(0001)\big)
=\widehat{Q}_n(0000)= Q_n(000).
 \end{equation}
Others, in general, will not simplify, and will have to be expressed by nested $\Upsilon$
functions, for example
\begin{multline}
 \widehat{Q}_n(00100)=
 \Upsilon\Big(\Upsilon\big(Q_n(001),Q_n(010),Q_n(011)\big),
              \Upsilon\big(Q_n(010),Q_n(100),Q_n(101)\big), \\
%&\mbox{\hspace{4em} }
\Upsilon_{0,1}\big(Q_n(010),Q_n(100),Q_n(101)\big)
\Big).
\end{multline}
Once we express all $\widehat{Q}_n(a_1a_2a_3a_4a_5)$ in eq. (\ref{rule26-k3}) by 3-block probabilities
$Q_n(a_1a_2a_3)$, we obtain
a map $[0,1]^8 \to [0,1]^8$. We omit explicit formulae for this map due to its complexity.
One should stress, however, that only four components of this map are independent, and
that by exploiting consistency conditions for block probabilities it is possible to
reduce this map to  $[0,1]^4 \to [0,1]^4$. We refer interested reader to \cite{paper50}, 
where we explained how to perform such reduction for local structure maps (the
same method can used for minimal entropy maps).

Just for the sake of comparison, let us also write local structure map of order
3 for rule~26. It can be obtained from eq. (\ref{rule26-k3}) by replacing $\widehat{Q}$
with $\widetilde{Q}$, 
\begin{align}\label{rule26-lst3}
Q_{n+1}(000)&=
\widetilde{Q}_n(00000)+
\widetilde{Q}_n(01010)+
\widetilde{Q}_n(10101)+
\widetilde{Q}_n(11010)+
\widetilde{Q}_n(11101)\nonumber \\
 &\mbox{\hspace{4em}}+
\widetilde{Q}_n(11110)+
\widetilde{Q}_n(11111), \nonumber \\
Q_{n+1}(001)&=
\widetilde{Q}_n(00001)+
\widetilde{Q}_n(01011)+
\widetilde{Q}_n(10100)+
\widetilde{Q}_n(11011)+
\widetilde{Q}_n(11100), \nonumber \\
Q_{n+1}(010)&=
\widetilde{Q}_n(00010)+
\widetilde{Q}_n(01000)+
\widetilde{Q}_n(10110)+
\widetilde{Q}_n(10111)+
\widetilde{Q}_n(11000), \nonumber \\
Q_{n+1}(011)&=
\widetilde{Q}_n(00011)+
\widetilde{Q}_n(01001)+
\widetilde{Q}_n(11001), \nonumber \\
Q_{n+1}(100)&=
\widetilde{Q}_n(00101)+
\widetilde{Q}_n(01101)+
\widetilde{Q}_n(01110)+
\widetilde{Q}_n(01111)+
\widetilde{Q}_n(10000), \nonumber \\
Q_{n+1}(101)&=
\widetilde{Q}_n(00100)+
\widetilde{Q}_n(01100)+
\widetilde{Q}_n(10001), \nonumber \\
Q_{n+1}(110)&=
\widetilde{Q}_n(00110)+
\widetilde{Q}_n(00111)+
\widetilde{Q}_n(10010), \nonumber \\
Q_{n+1}(111)&=
\widetilde{Q}_n(10011),
\end{align}
where
\begin{equation}
 \widetilde{Q}_n(a_1a_2a_3a_4a_5)=\frac{Q_n(a_1a_2a_3)Q_n(a_2a_3a_4)Q_n(a_3a_4a_5)}{
\big( Q_n(a_2a_30)+Q_n(a_2a_31) \big) \big(Q_n(a_3a_40)+Q_n(a_3a_41)\big)}.
\end{equation}

Both minimal entropy maps and local structure maps become rather complicated when $k$
increases. Because of high dimensionality and strong nonlinearity, it is difficult
to perform standard stability analysis for these maps.
It is, however, rather straightforward to write a computer program
which constructs and iterates them. 
 \section{Experimental results}
As we already mentioned, orbits of minimal entropy maps approximate 
orbits of measures under cellular automata rules. By iterating the minimal
entropy map, we can obtain approximate $P_n(\mathbf{a})$, that is,
probability of occurrence of block $\mathbf{a}$ after $n$ iterations
of a given cellular automata rule. How good is this approximation,
and it is any better than the local structure approximation?

In order to shed some light on  this question, we considered the
following problem.
Suppose that the initial measure is a Bernoulli measure $\mu_p$, so that
\begin{equation}\label{Bern}
\mu_p([\mathbf{a}])=P_0(\mathbf{a})=p^j(1-p)^{|\mathbf{a}|-j},
\end{equation}
where $j$ is the number
of ones in $\mathbf{a}$, $|\mathbf{a}|-j$ is the number of zeros in $\mathbf{a}$, and
$p \in [0,1]$. Probability of occurrence of $\mathbf{a}$ after $n$ iterations is then given by
\begin{equation}
 P_n(\mathbf{a})= (F^n\mu_p)([\mathbf{a}]).
\end{equation}
The expected value of a given cell after  $n$-th iteration of the rule, to be denoted $\rho_n$, is
given by
\begin{equation}
 \rho_n=1 \cdot P_n(1) + 0 \cdot P_n(0)=P_n(1).
\end{equation}
We will call $\rho_n$ a \emph{density of ones} at time $n$. Density can be estimated numerically
by starting with an array of $N$ sites and setting each one of them independently to 1 or 0
with probability $p$ or $1-p$, respectively. We then iterate rule $F$ $n$ times (using periodic boundary conditions)
 and count how many
cells are in state 1. The count divided by $N$ serves as a numerical estimate of $\rho_n$.

One can also estimate $\rho_n$ by iterating $k$-th order minimal entropy map $n$ times starting from
initial conditions given by eq. (\ref{Bern}), that is, $Q_0(\mathbf{a})=P_0(\mathbf{a})$. Then we compute $Q_n(1)$ by using
consistency conditions,
\begin{equation}
Q_n(1)=\sum_{\mathbf{a} \in \{0,1\}^{k-1}} Q_n(\mathbf{a}1),  
\end{equation}
and $Q_n(1)$ is used as an approximation of $\rho_n$,
to be called $k$-th order minimal entropy approximation of $\rho_n$.
Analogous approximation using local structure map will be called $k$-th order local structure
 approximation of $\rho_n$.

An interesting question is now how $\rho_n$ depends on $\rho_0$. Plot of $\rho_n$ vs. $\rho_0$ is
called \emph{density response curve}. We plotted density response curves using
``experimental'' $\rho_n$ as well as using minimal entropy approximation and
local structure approximation, both for orders $k=1,2,\ldots 7$.
 We found that, generally, as the order of the approximation
increases, density response curves obtained by iterating minimal entropy maps
become closer and closer to ``experimental curves''. The same phenomenon
is observed for density response curves obtained by iterating local structure
maps. 

For most elementary rules, both local structure maps and minimal entropy maps 
produce good approximations of density curves. There are two exceptions,
however, elementary CA rules 26 and 41. Here we will discuss rule 26 as an example.
\begin{figure}
 \begin{center}
   (a)\includegraphics[scale=1.1]{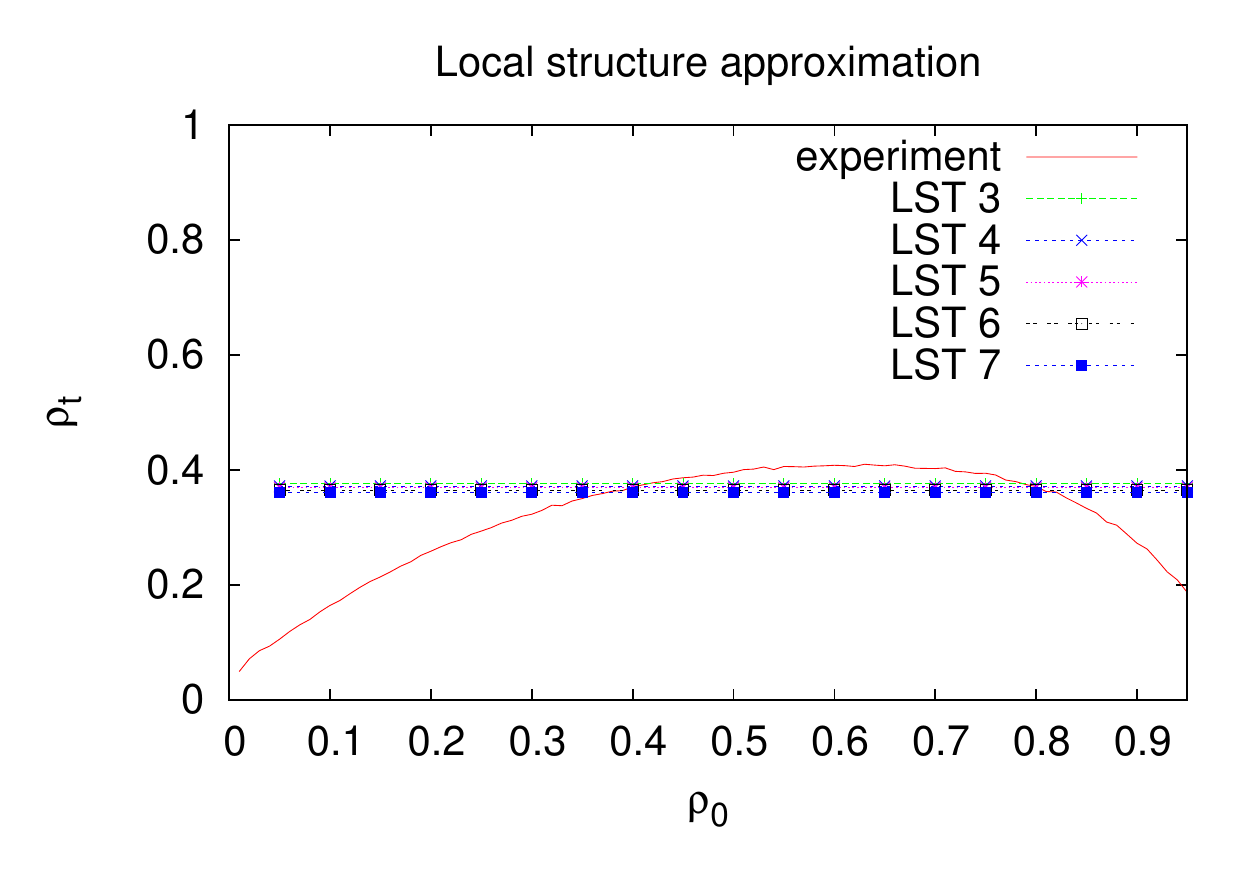} \\
   (b)\includegraphics[scale=1.1]{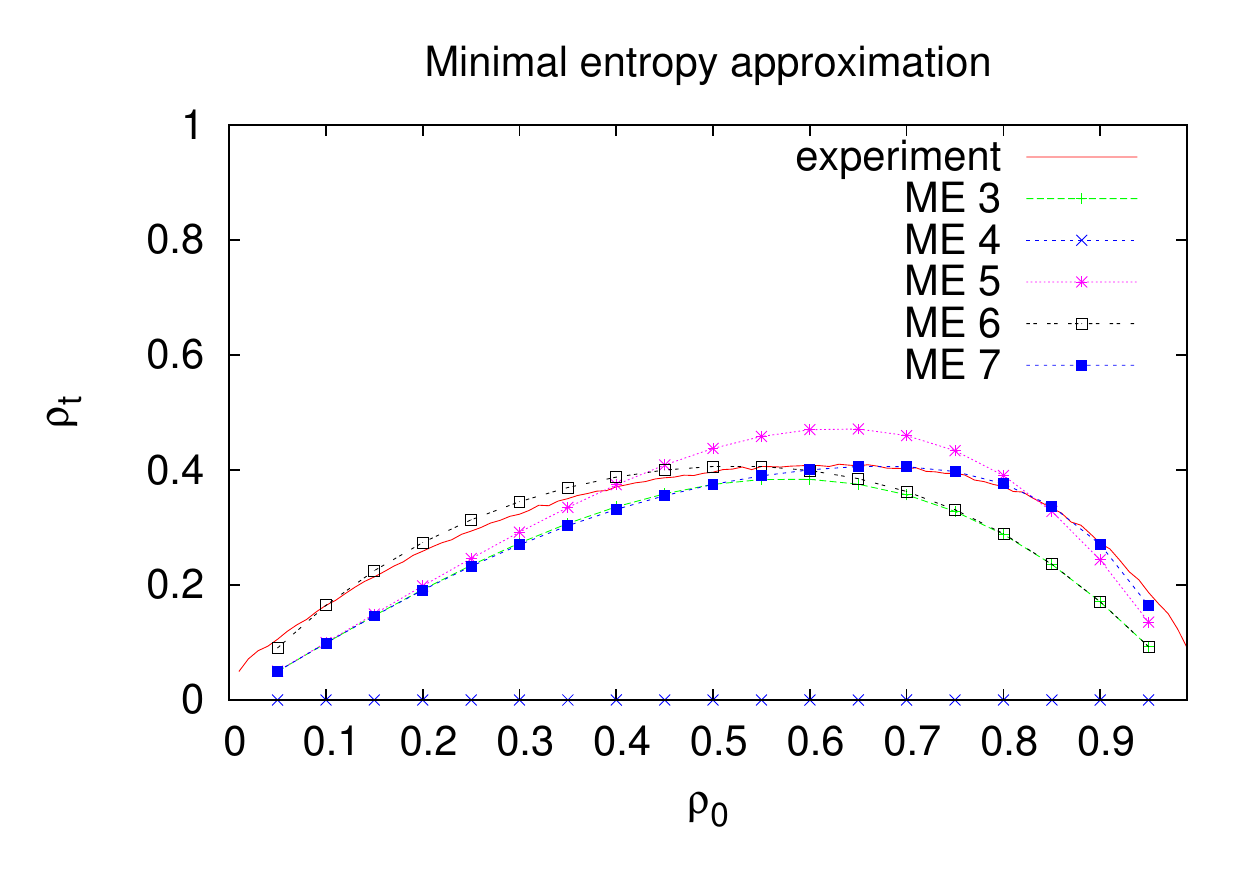}
   \end{center}
\caption{Density response curves for rule 26 for $t=10^3$ obtained
by iteration of local structure maps (a) and
minimal entropy maps (b).}\label{r26minent}
\end{figure}
The experimental density response curve is shown as the continuous curve in
Figure~\ref{r26minent}. Remarkably, density curves obtained by iterations of
local structure maps up to order 7 are horizontal straight lines, as shown
in Figure~\ref{r26minent}(a). One can say, therefore, that the local structure 
fails to predict the correct shape of the density curve, at least for $k\leq 7$.

In contrast to this, density curves obtained by iterations of
minimal entropy maps, shown in Figure~\ref{r26minent}(a), approximate the
shape of the ``experimental'' density curve much better, even at order 3.
The minimal entropy approximation, therefore, clearly outperforms the local
structure approximation in this case.
 
\section{Conclusions}
We introduced the notion of the minimal entropy approximation of probability
measures over binary bisequences. Minimal entropy approximation
can be viewed as an opposite of Bayesian approximation,
which maximizes entropy. We then demonstrated how the minimal entropy approximation 
can be used to construct approximations of orbits of measures
under the action of deterministic or probabilistic
cellular automata. Such approximate orbits can be fully
characterized by orbits of finite-dimensional maps,
which we call minimal entropy maps.
While points of approximate orbits of measures obtained by iterating
minimal entropy maps weakly converge to corresponding 
points of the exact orbits, just as in the case of   
approximate orbits of local structure theory, there are
cases when the minimal entropy approximation works
better than the local structure approximation. This 
is the case for elementary CA rule 26, for which the
local structure theory fails in predicting the correct shape of the
density response curve for $k \leq 7$. The minimal entropy approximation
yields fairly accurate prediction for the density response curve
of rule 26, starting with $k=3$.

An interesting question is  why is the minimal
entropy approximation better than the maximal entropy
approximation in the case of rule 26? One could naively 
think that this is because the time evolution of
rule 26 is somewhat more ``ordered'' than for other rules.
It is, however, not true: there are other rules for 
which the spatiotemporal patters are even more ``ordered''
than for rule 26, yet both maximal and minimal entropy
approximations seem to work for them equally well. 
In order to probe this issue further, one will need
to find more examples of rules for which the minimal
entropy approximation outperforms the local structure
theory. A natural way  to go beyond elementary CA rules
considered here is to search for such examples among either probabilistic
CA rules of radius 1, or deterministic CA rules of radius
grater than 1.  Both possibilities are currently
investigated by the author.

\section{Acknowledgements}
The author acknowledges partial financial support from the Natural
Sciences and Engineering Research Council of Canada (NSERC) in the
form of Discovery Grant. Some calculations on which this work is based were made
 possible by the facilities of the Shared
Hierarchical Academic Research Computing Network (SHARCNET:www.sharcnet.ca) and
Compute/Calcul Canada. 

\providecommand{\href}[2]{#2}\begingroup\raggedright\endgroup
\end{document}